\documentclass[aps,pra,reprint,superscriptaddress]{revtex4-1}
\usepackage{amsmath}
\usepackage{amsfonts}
\usepackage{hyperref}
\usepackage{graphicx}

\begin{document}
\title{The multi-mode quantum Entropy Power Inequality}
\author{G. De Palma}
\affiliation{NEST, Scuola Normale Superiore and Istituto Nanoscienze-CNR, I-56127 Pisa,
Italy.}
\affiliation{INFN, Pisa, Italy}
\author{A. Mari}
\affiliation{NEST, Scuola Normale Superiore and Istituto Nanoscienze-CNR, I-56127 Pisa,
Italy.}
\author{S. Lloyd}
\affiliation{Research Laboratory of Electronics, Massachusetts Institute of Technology, Cambridge, MA 02139, USA.}
\affiliation{Department of Mechanical Engineering, Massachusetts Institute of Technology, Cambridge, MA 02139, USA.}
\author{V. Giovannetti}
\affiliation{NEST, Scuola Normale Superiore and Istituto Nanoscienze-CNR, I-56127 Pisa,
Italy.}

\begin{abstract}
The quantum version of a fundamental  entropic data-processing  inequality is presented. It establishes a lower bound for the entropy which can be generated in the output channels of a scattering process which involves a collection of  independent input bosonic modes (e.g. the modes of the electromagnetic field). The impact of this inequality in quantum information theory is potentially large and some relevant implications are considered in this work.
\end{abstract}

\maketitle

\section{Introduction}
\begin{figure}[t]
\includegraphics[width=1 \columnwidth]{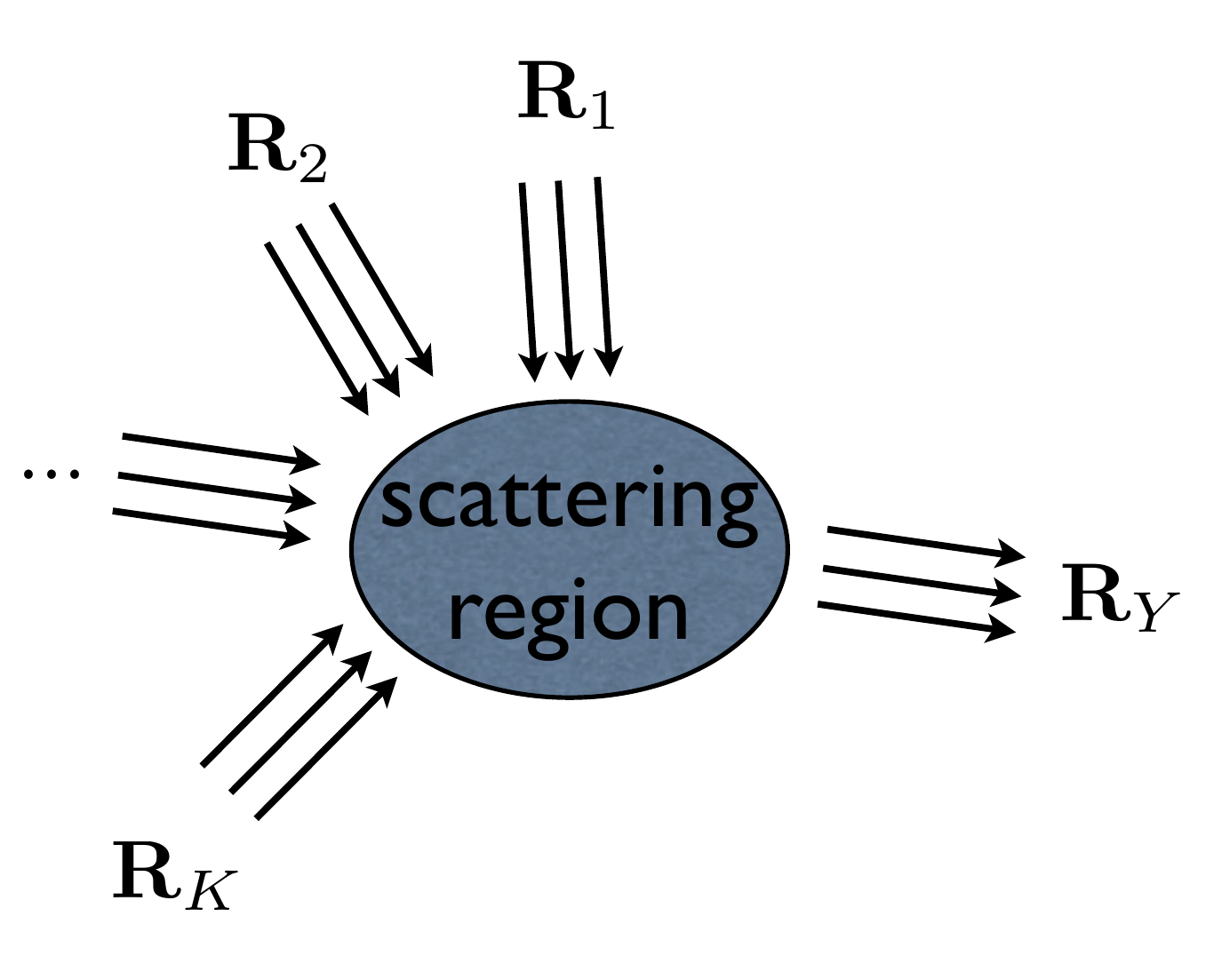}
\caption{(Color online): Graphical representation of the scheme underlying the multi-mode qEPI~(\ref{EPIrate}): it establishes a lower bound on the von Neumann entropy  emerging from the output port indicated by $\mathbf{R}_Y$ of a multi-mode scattering process that linearly couples
$K$ independent sets of bosonic input modes (each containing $n$ modes), initialized into factorized density matrices. }\label{model}
\end{figure}

Entropic inequalities are a fundamental tool in classical information and communication theory \cite{coverthomas}, where they can be used  to bound the efficiency of data processing procedures. For this reason, a large effort has been devoted to this subject, with results such as the Entropy Power Inequality \cite{Shannon,stam,blachman,verdu,rioul,guo}, used in the proof of a stronger version of the central limit theorem \cite{CLT} and crucial in the computation of the capacities of various classical channels~\cite{bergmans}, and the Brunn-Minkowski inequality (for a review, see \cite{dembo} or chapter 17 of \cite{coverthomas}). For the same reason, entropic inequalities are fundamental also in the context of quantum information theory~\cite{BENSHOR}.
 In particular the long-standing problem of determining  the classical capacity of phase-insensitive quantum bosonic Gaussian channels~\cite{HOWE,weedbrook}  was  linked to a lower bound conjectured to hold for the minimum von Neumann entropy achievable at the output of a transmission line (the Minimum Output Entropy conjecture, MOE)~\cite{MINCON}.
 While these issues were recently solved in Refs.~\cite{ghgp,ggpch,mgh}
a stronger version of the MOE relation, arising  from a suitable quantum generalization of the Entropy Power Inequality, is still not proved.
This new relation, called Entropy Photon-number Inequality (EPnI)~\cite{EPnIguha},  turns out to be crucial  in the determining the classical capacity regions of the quantum bosonic broadcast \cite{guhaieee,guhabroadcast} and wiretap \cite{guha} channels. A partial solution  has been provided in~\cite{dPMG} by proving
a weaker version of the EPnI,  called quantum Entropy Power Inequality (qEPI) and
  first introduced and studied by K\"onig and Smith in Ref.~\cite{ks,ks-natphot}.
Both the EPnI and  the qEPI establish lower bounds on the entropy achievable in one of the output channels  originating when two bosonic input modes, initialized in factorized input states of assigned entropies, are coupled via  a beam-splitter or an amplifier transformation~\cite{WALLS}.
Here we present a multi-mode generalization of the qEPI which applies to the context
where an arbitrary collection of independent input bosonic modes undergo to a scattering process which mixes them according to some linear coupling - see Fig.~\ref{model} for a schematic representation of the model.
This new inequality permits to put bounds on the MOE inequality, still unproved for non gauge-covariant multi-mode channels, and then on the classical capacity of any (not necessarily phase-insensitive) quantum Gaussian channel.
Besides, our finding  can find potential applications in extending the results of \cite{dPMG} on the classical capacity region of the quantum bosonic broadcast channel to the Multiple-Input Multiple-Output setting (see e.g. Ref.~\cite{CAVES}), providing upper bounds for the associated capacity regions.

\section{The problem}
The generalization of the qEPI we discuss in the present work finds a classical analogous in the multi-variable version of the EPI~\cite{Shannon,stam,blachman,verdu,rioul,guo}. The latter applies to a set
of $K$ independent random variables $\mathbf{X}_\alpha,\;\alpha=1,\ldots,K$, valued in $\mathbb{R}^m$ and collectively denoted by $\mathbf{X}$, with factorized probability densities $p_X(\mathbf{x})=p_1(\mathbf{x}_1)\ldots p_K(\mathbf{x}_K)$, and with Shannon differential entropies \cite{Shannon} $H_\alpha=-\left\langle\ln p_\alpha(\mathbf{x}_\alpha)\right\rangle$ (the $\langle \cdots \rangle$ representing the average with respect to the associated probability distribution).
Defining hence the linear combination
\begin{equation}\label{Ycl}
\mathbf{Y}=M\,\mathbf{X}=\sum_{\alpha=1}^K M_{\alpha}\,\mathbf{X}_\alpha\;,
\end{equation}
where $M$ is an $m\times Km$ real matrix made by the $K$ blocks $M_\alpha$, each of dimension $m\times m$, the multi-variable EPI gives an (optimal) lower bound to the Shannon entropy $H_Y$ of $\mathbf{Y}$
\begin{equation}\label{cEPI}
\exp[{2}H_Y/m]\geq\sum_{\alpha=1}^K|\det M_\alpha|^\frac{2}{m}\;\exp[{2}H_\alpha/m]\;.
\end{equation}
In the original derivation~\cite{Shannon,stam,blachman,verdu,rioul,guo} this inequality  is proved under the assumption that
all the $M_\alpha$ coincide with the identity matrix, i.e. for
$\mathbf{Y}=\sum_{\alpha=1}^K\widetilde{\mathbf{X}}_\alpha$.
From this however Eq.~\eqref{cEPI} can be easily established choosing $\widetilde{\mathbf{X}}_\alpha=M_\alpha\mathbf{X}_\alpha$, and remembering that the entropy $\widetilde{H}_\alpha$ of $\widetilde{\mathbf{X}}_\alpha$ satisfies $\widetilde{H}_\alpha=H_\alpha+\ln|\det M_\alpha|$.
It is also worth observing that for Gaussian variables the exponentials of
the entropies $H_\alpha$ and $H_\mathbf{Y}$
 are proportional to the determinant of the corresponding  covariance matrices, i.e.
 $$H_\alpha=\frac{1}{2}\ln\det\left(\pi e\,\sigma_\alpha\right)$$
and
$$H_\mathbf{Y}=\frac{1}{2}\ln\det\left(\pi e\,\sigma_\mathbf{Y}\right)\;,$$
with
$$\sigma_\alpha=2\left\langle\Delta \mathbf{x}_\alpha \,\Delta \mathbf{x}_\alpha^T\right\rangle\;,\qquad\sigma_Y=2\left\langle\Delta \mathbf{y} \,\Delta \mathbf{y}^T\right\rangle$$
and
$$\Delta \mathbf{x}_\alpha = \mathbf{x}_\alpha - \left\langle \mathbf{x}_\alpha \right\rangle\;,\qquad\Delta \mathbf{y} = \mathbf{y} - \left\langle \mathbf{y} \right\rangle\;.$$
 Accordingly  in this special case  Eq.~(\ref{cEPI}) can be seen as
an  instance  of the Minkowski's determinant inequality~\cite{INEQU} applied to the identity
\begin{equation}\label{cmphi}
\sigma_Y=\sum_{\alpha=1}^KM_\alpha\,\sigma_\alpha\,M_\alpha^T\;,
\end{equation}
  and it saturates
under the assumption that the matrices entering the sum are  all proportional to a given matrix $\sigma$, i.e.
\begin{eqnarray} \label{condition}
M_\alpha\,\sigma_\alpha\,M_\alpha^T = c_\alpha\,\sigma\;,
\end{eqnarray}
with $c_\alpha$ being arbitrary (real) coefficients.

In the quantum setting the random variables get replaced by $n=\frac{m}{2}$ bosonic modes (for each mode there are two quadratures, $Q$ and $P$), and instead of probability distributions over $\mathbb{R}^{2n}$, we have the quantum density matrices $\hat{\rho}_\alpha$ on the Hilbert space $L^2(\mathbb{R}^n)$.
For each $\alpha$, let $\mathbf{\hat{R}}_\alpha$ be the column vector that collectively denotes all the quadratures of the $\alpha$-th subsystem:
\begin{equation}
\mathbf{\hat{R}}_\alpha=\left(\hat{Q}_\alpha^1,\;\hat{P}_\alpha^1,\ldots,\;\hat{Q}_\alpha^n,\;\hat{P}_\alpha^n\right)^T\;,\quad\alpha=1,\ldots,\,K\;.
\end{equation}
The $\mathbf{\hat{R}}_\alpha$ satisfy the canonical commutation relations
\begin{equation}\label{CCR}
\left[\mathbf{\hat{R}}_\alpha,\;\mathbf{\hat{R}}_\beta^T\right]=\delta_{\alpha\beta}\;\Delta\;\hat{\openone}\;,
\end{equation}
where $\Delta$ is the symplectic matrix  (see e.g. \cite{channels}) given by
$$\Delta=\bigoplus_{k=1}^n\left(
                                                                                                        \begin{array}{cc}
                                                                                                          0 & 1 \\
                                                                                                          -1 & 0 \\
                                                                                                        \end{array}
                                                                                                      \right)\;.$$
Consider then totally factorized input states $\hat{\rho}_X=\bigotimes_{\alpha=1}^K\hat{\rho}_\alpha$, where $\hat{\rho}_\alpha$ is the density matrix of the $\alpha$-th input.
The analogue of \eqref{Ycl} is defined with
\begin{equation}\label{channel}
\hat{\rho}_Y=\Phi\left(\hat{\rho}_X\right)=\mathrm{Tr}_Z\left(\hat{U}\;\hat{\rho}_X\;\hat{U}^\dag\right)\;,
\end{equation}
where $\hat{U}:\mathcal{H}_X\longrightarrow\mathcal{H}_Y\otimes\mathcal{H}_Z$ is an isometry between the input Hilbert space $\mathcal{H}_X$ and the tensor product of the output Hilbert space $\mathcal{H}_Y$ with an ancilla Hilbert space $\mathcal{H}_Z$, satisfying
\begin{equation}\label{Yq}
\hat{U}^\dag\;\mathbf{\hat{R}}_Y\;\hat{U}=M\,\mathbf{\hat{R}}_X=\sum_{\alpha=1}^K M_{\alpha}\,\mathbf{\hat{R}}_\alpha\;.
\end{equation}
As before, $M$ is a $2n\times 2Kn$ real matrix made by the $2n\times 2n$ square blocks $M_\alpha$.
The canonical commutation relations \eqref{CCR} on $\mathbf{\hat{R}}_Y$ together with the unitarity of $\hat{U}$ impose the constraint
$$\sum_{\alpha=1}^K M_\alpha\Delta M_\alpha^T=\Delta\;.$$
Notice that at the level of the covariance matrices  Eq.~(\ref{Yq}) induces the same mapping~(\ref{cmphi}) that holds in the classical scenario
(in this case however  one has
$$\sigma_\alpha:=\left\langle\left\{\mathbf{\hat{R}}_\alpha-\left\langle\mathbf{\hat{R}}_\alpha\right\rangle,\;\mathbf{\hat{R}}_\alpha^T-\left\langle\mathbf{\hat{R}}_\alpha^T\right\rangle\right\}\right\rangle\;,$$ $$\sigma_Y:=\left\langle\left\{\mathbf{\hat{R}}_Y-\left\langle\mathbf{\hat{R}}_Y\right\rangle,\;\mathbf{\hat{R}}_Y^T-\left\langle\mathbf{\hat{R}}_Y^T\right\rangle\right\}\right\rangle$$ with
 $\langle \cdots \rangle = \mathrm{Tr}[\hat{\rho}_X \cdots ]$ and $\{ \cdots, \cdots\}$ representing the anti-commutator).
The isometry $\hat{U}$ in \eqref{channel} does not necessarily conserve energy, i.e. it can contain active elements, so that even if the input $\hat{\rho}_X$ is the vacuum on all its $K$ modes, the output $\hat{\rho}_Y$ can be thermal with a nonzero temperature.
For $K=2$, the beam-splitter of parameter $0\leq\lambda\leq1$ is easily recovered with
$$M_1=\sqrt{\lambda}\;\openone_{2n}\;,\qquad M_2=\sqrt{1-\lambda}\;\openone_{2n}\;.$$
To get the quantum amplifier of parameter $\kappa\geq1$, we must take instead
$$M_1=\sqrt{\kappa}\;\openone_{2n}\;,\qquad M_2=\sqrt{\kappa-1}\;T_{2n}\;,$$
where $T_{2n}$ is the $n$-mode time-reversal
$$T_{2n}=\bigoplus_{k=1}^n\left(
                            \begin{array}{cc}
                              1 & 0 \\
                              0 & -1 \\
                            \end{array}
                          \right)\;.$$

We can now state the multi-mode qEPI: the von Neumann entropies $S_\alpha=-\mathrm{Tr}\left(\hat{\rho}_\alpha\ln\hat{\rho}_\alpha\right)$ satisfy the analogue of \eqref{cEPI}
\begin{equation}
\exp[{S_Y}/{n}]\geq\sum_{\alpha=1}^K \lambda_\alpha\;\exp[{S_\alpha}/{n}]\;,\label{EPI}
\end{equation}
where we have defined $\lambda_\alpha:=\left|\det M_\alpha\right|^\frac{1}{n}$.
The qEPI \eqref{EPI} was proved \cite{ks,dPMG} only in the simple cases of the quantum beam-splitter and amplifier. As already noticed,  in the classical setting the generalized inequality \eqref{cEPI} is a trivial consequence of the case with all the $M_\alpha$ equal to the identity. In the quantum setting this is not the case and one needs to find a proof that works for all
possible choices of  the $M_\alpha$. The main result of the present paper is exactly to tackle this problem.

\section{The proof}
The proof of Eq.~(\ref{EPI}), even if with some non trivial modifications, proceeds along the same line of  the one in \cite{ks} and \cite{dPMG}.
Specifically, inspired from what we know about the classical case,
 we expect that the  qEPI should be saturated by quantum Gaussian states~\cite{weedbrook,channels}  with high entropy and whose  covariance matrices $\sigma_\alpha$
 fulfill the condition~(\ref{condition})
 (the high entropy limit being necessary to ensure that the associated quantum Gaussian states behave  as classical Gaussian probability distributions).
Suppose hence we do apply a transformation on the input modes of the system which depends on a real parameter $\tau$ that plays the role of an effective temporal coordinate,  and which is constructed in  such a way that, starting from $\tau=0$ from the input state $\hat{\rho}_X$  it will drive the modes towards such optimal  Gaussian configurations
 in the asymptotic limit  $\tau\rightarrow\infty$ --- see Sec.~\ref{evol}.
 Accordingly for each $\tau\geq 0$ we will have an associated value for the entropies $S_\alpha$ and $S_Y$ which, if the qEPI is correct, should still fulfill the bound~(\ref{EPI}).
To verify this it is useful to put the qEPI \eqref{EPI} in the rate form
\begin{equation}
\frac{\sum_{\alpha=1}^K \lambda_\alpha\;\exp[{S_\alpha}/{n}]}{\exp[{S_Y}/{n}]}\leq1\label{EPIrate}\;.
\end{equation}
We will then study  the  left-hand-side of Eq.~\eqref{EPIrate}  showing that its
parametric  derivative is always positive  (see Sec.~\ref{sec:para}) and that that for $\tau\rightarrow \infty$ it tends to 1  (see section \ref{appscaling}).

\subsection{Parametric evolution}\label{evol}

In this section we find the parametric evolution suitable to the proof.
 For this purpose  for each input mode $\alpha$ we  enforce
the following dynamical process
 \begin{eqnarray}
\frac{d}{dt}\hat{\rho}_\alpha(t) =\mathcal{L}_{\gamma_\alpha}(\hat{\rho}_\alpha(t))\;,\label{liouville11}
\end{eqnarray}
characterized by  the Lindblad super-operator
\begin{equation}
\mathcal{L}_{\gamma_\alpha}(\hat{\rho})\label{liouville}:=-\frac{1}{4}\left[\left(\Delta^{-1}\mathbf{\hat{R}}_\alpha\right)^T,\;\gamma_\alpha\left[\Delta^{-1}\mathbf{\hat{R}}_\alpha,\;\hat{\rho}\right]\right]\;,
\end{equation}
where if $M_\alpha$ is invertible, $\gamma_\alpha:=\lambda_\alpha M_\alpha^{-1}M_\alpha^{-T}$ is positive definite, and if $M_\alpha$ is not invertible, $\gamma_\alpha:=0$, i.e. we do not evolve the corresponding input.
The generator~\eqref{liouville} commutes with translations, i.e.
\begin{equation}
\mathcal{L}_{\gamma_\alpha}\left(\hat{D}_\mathbf{x}\;\hat{\rho}\; \hat{D}_\mathbf{x}^\dag\right)=\hat{D}_\mathbf{x}\;\mathcal{L}_{\gamma_\alpha}(\hat{\rho})\;\hat{D}_\mathbf{x}^\dag\;,\label{liouvtr}
\end{equation}
where $\hat{D}_\mathbf{x}:=\exp\left(i\mathbf{x}^T\Delta^{-1}\mathbf{\hat{R}}_\alpha\right)$ are  the displacement operators of the system~\cite{channels}. Furthermore
it induces a diffusive evolution which   adds Gaussian noise into the system driving it  toward the set of Gaussian states while inducing a linear increase of the mode covariance matrix, i.e. \begin{eqnarray}
\frac{d}{dt} \sigma_\alpha(t) =   \gamma_\alpha  \quad \Longrightarrow \quad
\sigma_\alpha(t) =  \sigma_\alpha(0) + t\,\gamma_\alpha\;,
\label{sigma} \end{eqnarray} which boosts its entropy. Notice that the choice we made on $\gamma_\alpha$ ensures that for large enough $t$, $M_\alpha \sigma_\alpha(t)M_\alpha^T$ will asymptotically approach the saturation condition~(\ref{condition}) of the classical EPI with the matrix $\sigma$ being the identity operator.
 We now let the various input modes  evolve independently with their own
 processes~(\ref{liouville11})  for different time intervals $t_\alpha\geq 0$:  accordingly
the input state of the system is  mapped from $\hat{\rho}_X$ to $\hat{\rho}_X(t_1,t_2, \cdots, t_K) =\bigotimes_{\alpha=1}^K\hat{\rho}_\alpha(t_\alpha)$
with $\hat{\rho}_\alpha(t_\alpha)$ being the evolved density matrix of the $\alpha$-mode, its  von Neumann entropy being $S_\alpha(t_\alpha)$.
Next in order to get a one parameter trajectory  we link the various time intervals $t_\alpha$ by
parametrizing them in terms of an external coordinate $\tau\geq 0$ by enforcing  the following constraint:
\begin{eqnarray} \label{choice}
\frac{d}{d\tau}t_\alpha(\tau)=\exp[S_\alpha(t_\alpha(\tau))/n]\;, \qquad t_\alpha(0)=0\;.
\end{eqnarray}
This is a first order differential equation which, independently from the particular functional dependence of  $S_\alpha(t_\alpha)$, always admits a solution. Furthermore,
since the right-hand-side of Eq.~(\ref{choice}) is always greater than or equal to 1, it follows that $t_\alpha(\tau)$ diverges as $\tau$ increases, i.e.
\begin{eqnarray} \label{div} \lim_{\tau \rightarrow \infty} t_\alpha(\tau) = \infty\;.\end{eqnarray}
Accordingly in the asymptotic limit
of large $\tau$, the mapping   $\hat{\rho}_X\rightarrow \hat{\rho}_X(\tau) =\bigotimes_{\alpha=1}^K\hat{\rho}_\alpha(t_\alpha(\tau))$  will drive the system toward the tensor product of the asymptotic points defined by the diffusive local master equation~(\ref{liouville11}). As we shall see in Sec.~\ref{appscaling} this implies that the rate on the left-hand-side of Eq.~(\ref{EPIrate}) will asymptotically reach the value $1$. In order to evaluate this limit, as well as to study the parametric derivative in $\tau$ of such a rate, we need to compute
the functional dependence upon $\tau$ of the von Neumann entropy  $S_Y$ of the output modes associated with the coordinates
$\mathbf{\hat{R}}_Y$. It turns out that with the choice~(\ref{choice}), the parametric evolution of the input mode induces a temporal
evolution  of the output modes which, expressed in terms of the local time coordinate $t_Y$ having parametric dependence upon $\tau$ given by
\begin{equation}\label{tY}
t_Y(\tau)=\sum_{\alpha=1}^K\lambda_\alpha t_\alpha(\tau)\;,
\end{equation}
 is still in the form of (\ref{liouville11}) with the operators
$\mathbf{\hat{R}}_\alpha$ appearing in~(\ref{liouville}) being replaced by $\mathbf{\hat{R}}_Y$, and with the matrix $\gamma_\alpha$  being replaced by
$\openone_{2n}$.
Accordingly in this case Eq.~(\ref{sigma}) becomes
\begin{equation}
\frac{d\sigma_Y}{d t_Y}=  \openone_{2n} \;\; \Longrightarrow \;\;
\sigma_Y(t_Y)= \sigma_Y(0) +  t_Y\,\openone_{2n}\label{sigmay}\;.
\end{equation}

\subsection{Evaluating the parametric derivative of the rate
}\label{sec:para}
Define the Fisher information matrix of a quantum state $\hat{\rho}$ as the Hessian with respect to $\mathbf{x}$ of the relative entropy between $\hat{\rho}$ and $\hat{\rho}$ displaced by $\mathbf{x}$:
\begin{equation}
J_{ij}(\hat{\rho}):=\left.\frac{\partial^2}{\partial x^i\partial x^j}S\left(\hat{\rho}\left\|\hat{D}_\mathbf{x}\hat{\rho}\hat{D}_\mathbf{x}^\dag\right.\right)\right|_{\mathbf{x}=0}\;.
\end{equation}
An explicit computation shows
\begin{equation}
J=\mathrm{Tr}\left(\left[\Delta^{-1}\mathbf{\hat{R}}\,,\,\left[\left(\Delta^{-1}\mathbf{\hat{R}}\right)^T,\;\hat{\rho}\right]\right]\ln \hat{\rho}\right)\label{Jij}\;.
\end{equation}
The key observation is that the Fisher information matrix is easily related to the derivative of the entropy with respect to the evolution \eqref{liouville} through a generalization of the de Bruijn identity of \cite{ks,dPMG}:
\begin{equation}
\frac{d}{dt}S\left(\hat{\rho}(t)\right)=\frac{1}{4}\mathrm{tr}\left(J\left(\hat{\rho}(t)\right)\;\frac{d}{dt}\sigma(t)\right)\;.\label{debrujin}
\end{equation}
With \eqref{debrujin} and \eqref{choice} we can compute the time derivatives of the entropies  which enter in the definition of the rate  in the left-hand-side
of Eq.~(\ref{EPIrate}). Specifically from Eqs.~(\ref{sigma}), (\ref{choice}),  (\ref{tY}) and (\ref{sigmay}) we get
\begin{eqnarray}
\frac{dS_\alpha}{d\tau}&=&\frac{1}{4}e^{\frac{1}{n}S_\alpha}\mathrm{Tr}\left(J_\alpha\;\gamma_\alpha\right)\label{dsalpha}\\
\frac{dS_Y}{d\tau}&=&\frac{1}{4}\left(\sum_{\alpha=1}^Ke^{\frac{1}{n}S_\alpha}\lambda_\alpha\right)\mathrm{Tr}J_Y\;,\label{dsy}
\end{eqnarray}
where $J_Y$ and $J_\alpha$ are the quantum Fisher information matrices of the output and the $\alpha$-th input, respectively.

The next step is to exploit
the data processing inequality for the relative entropy between the state  $\hat{\rho}_X$ of the input modes and its displaced version $\hat{D}_{\mathbf{x}}\hat{\rho}_X\hat{D}_{\mathbf{x}}^\dag$, i.e.
\begin{equation}
S\left(\Phi\left(\hat{\rho}_X\right)\left\|\Phi\left(\hat{D}_{\mathbf{x}}\hat{\rho}_X\hat{D}_{\mathbf{x}}^\dag\right)\right.\right)\leq S\left(\hat{\rho}_X\left\|\hat{D}_{\mathbf{x}}\hat{\rho}_X\hat{D}_{\mathbf{x}}^\dag\right.\right)\;,\label{dataproc2}
\end{equation}
where $\Phi$ is the CPTP map defined in \eqref{channel}.
Another characterization of the latter can be obtained by exploiting the
 characteristic function representation of a quantum state $\hat{\rho}$,
 \cite{channels}
\begin{equation}\label{chi}
\chi(\mathbf{k}):=\mathrm{Tr}\left(\hat{\rho}\,e^{i\mathbf{k}^T\mathbf{\hat{R}}}\right)\;,\qquad\hat{\rho}=\int\chi(\mathbf{k})\;e^{-i\mathbf{k}^T\mathbf{\hat{R}}}\;\frac{d\mathbf{k}}{(2\pi)^n}\;.
\end{equation}
Let then $\chi_X\left(\mathbf{k}\right)$, $\mathbf{k}\in\mathbb{R}^{2Kn}$, and $\chi_Y\left(\mathbf{q}\right)$, $\mathbf{q}\in\mathbb{R}^{2n}$ be the characteristic functions of the input and the output, respectively. From \eqref{channel} and \eqref{Yq} we get
\begin{equation}\label{Phichi}
\chi_Y\left(\mathbf{q}\right)=\chi_X\left(M^T\mathbf{q}\right)\;,
\end{equation}
with $M$ being the matrix entering in Eq.~(\ref{Yq}).

We then notice that
displacing the inputs by $\mathbf{x}$, the output gets translated by
$$\mathbf{y}=M\mathbf{x}=\sum_{\alpha=1}^K M_\alpha \mathbf{x}_\alpha\;,$$
i.e.
$$\hat{D}_{\mathbf{y}}\Phi\left(\hat{\rho}_X\right)\hat{D}_{\mathbf{y}}^\dag=\Phi\left(\hat{D}_{\mathbf{x}}\hat{\rho}_X\hat{D}_{\mathbf{x}}^\dag\right)\;.$$
Therefore from (\ref{dataproc2}) it follows
\begin{eqnarray}
&&S\left(\Phi\left(\hat{\rho}_X\right)\left\|\hat{D}_{\mathbf{y}}\Phi\left(\hat{\rho}_X\right)\hat{D}_{\mathbf{y}}^\dag\right.\right) \leq S\left(\hat{\rho}_X\left\|\hat{D}_{\mathbf{x}}\hat{\rho}_X\hat{D}_{\mathbf{x}}^\dag\right.\right)=\nonumber\\
 &&=\sum_{\alpha=1}^K S\left(\hat{\rho}_\alpha\left\|\hat{D}_{\mathbf{x_\alpha}}\hat{\rho}_\alpha\hat{D}_{\mathbf{x_\alpha}}^\dag\right.\right)\;,
\label{dataproc}
\end{eqnarray}
where in the last passage we used the  additivity of the relative entropy on product states.
Since both the first and the last member of \eqref{dataproc} are nonnegative and vanishing for $\mathbf{x}=0$, inequality \eqref{dataproc} translates to the Hessians.
The variables are the $x^i_\alpha,\;i=1,\ldots,\,2n,\;\alpha=1,\ldots,\,K,$ so the Hessian is a matrix with indices $(i,\alpha),(j,\beta)$, and the inequality reads
\begin{equation}
\left(M_\alpha^TJ_YM_\beta\right)_{\alpha\beta}\leq\left(\delta_{\alpha\beta}J_\alpha\right)_{\alpha\beta}\;,\label{dataprocJ}
\end{equation}
where the indices $i,\,j$ are left implicit.
Finally, sandwiching \eqref{dataprocJ} with $\lambda_\alpha e^{\frac{1}{n}S_\alpha}M_\alpha^{-T}$ on the left and its transpose $\lambda_\beta e^{\frac{1}{n}S_\beta}M_\beta^{-1}$ on the right (if $M_\alpha$ is not invertible, $\lambda_\alpha=0$ and the corresponding terms are supposed to vanish), we get
\begin{equation}
\left(\sum_{\alpha=1}^K \lambda_\alpha\;e^{\frac{1}{n}S_\alpha}\right)^2\mathrm{tr}J_Y\leq\sum_{\alpha=1}^K \lambda_\alpha\; e^{\frac{2}{n}S_\alpha}\;\mathrm{tr}\left(J_\alpha\;\gamma_\alpha\right)\;,\label{stam}
\end{equation}
and computing the parametric derivative of the rate \eqref{EPIrate} with \eqref{dsalpha} and \eqref{dsy}, it is easy to show that \eqref{stam} is equivalent to its positivity.

\subsection{Asymptotic scaling}\label{appscaling}
In this section we prove that the rate \eqref{EPIrate} tends to one for $\tau\to\infty$. For this purpose, we need the asymptotic scaling of the entropy under the dissipative  evolution
described by Eqs.~\eqref{liouville11}, \eqref{liouville}. Remember that we are evolving only the inputs with invertible $M_\alpha$, for which $\gamma_\alpha>0$.

\subsubsection{A lower bound for the entropy}
A lower bound for the entropy follows on expressing the state $\hat{\rho}$ in terms of its generalized  Husimi function $Q_{\Gamma}(\mathbf{x})$, see e.g. Ref.~\cite{MAJ1}.
Specifically, given  a Gaussian state $\hat{\rho}_{\mathbf{x},\Gamma}$ characterized by first momentum $\mathbf{x}\in\mathbb{R}^{2n}$ and
covariance matrix $\Gamma\geq\pm i\Delta$,  we define
\begin{equation}\label{genHusimi}
Q_{\Gamma}(\mathbf{x}) := \frac{\mathrm{Tr}\left(\hat{\rho}\;\hat{\rho}_{\mathbf{x},\Gamma}\right)}{(2\pi)^n} = \int e^{-\frac{1}{4}\mathbf{k}^T\Gamma\mathbf{k}-i\mathbf{k}^T\mathbf{x}}\;\chi(\mathbf{k})\;\frac{d\mathbf{k}}{(2\pi)^{2n}}\;,
\end{equation}
 where in the second line we used (\ref{chi}) and the fact that  $e^{-\frac{1}{4}\mathbf{k}^T\Gamma\mathbf{k}+i\mathbf{k}^T\mathbf{x}}$ is the
 characteristic function of $\hat{\rho}_{\mathbf{x},\Gamma}$ (the conventional Husimi distribution~\cite{WALLS}  being recovered taking the states $\hat{\rho}_{\mathbf{x},\Gamma}$  to
 be displaced vacua, i.e. coherent states). By construction, $Q_{\Gamma}(\mathbf{x})$ is continuous in $\mathbf{x}$ and positive: $Q_{\Gamma}(\mathbf{x})\geq0$.
Furthermore  since $\chi(0)=\mathrm{Tr}\hat{\rho}=1$ for any normalized state $\hat{\rho}$, we also have
$$\int Q_{\Gamma}(\mathbf{x})\;d\mathbf{x}=1\;:$$
the generalized Husimi function $Q_{\Gamma}(\mathbf{x})$ is hence a probability distribution.
Taking the Fourier transform of  $Q_\Gamma(\mathbf{x})$,  Eq.~(\ref{genHusimi}) can now be inverted obtaining
\begin{equation}
\hat{\rho}=\int Q_\Gamma(\mathbf{x})\left(\int e^{\frac{1}{4}\mathbf{k}^T\Gamma\mathbf{k}+i\mathbf{k}^T\mathbf{x}}\;e^{-i\mathbf{k}^T\mathbf{\hat{R}}}\;\frac{d\mathbf{k}}{(2\pi)^n}\right)d\mathbf{x}\;.\label{rhoQ}
\end{equation}
Comparing with \eqref{chi} the integral in parenthesis, it looks like a Gaussian ``state'' with covariance matrix $-\Gamma$ displaced by $\mathbf{x}$. Of course, this is not a well-defined object, and it makes sense only if integrated against smooth functions as $Q_\Gamma(\mathbf{x})$. However, if we formally define
$$\hat{\rho}_{-\Gamma}:=\int e^{\frac{1}{4}\mathbf{k}^T\Gamma\mathbf{k}}\;e^{-i\mathbf{k}^T\mathbf{\hat{R}}}\;\frac{d\mathbf{k}}{(2\pi)^n}\;,$$
Eq.~\eqref{rhoQ} can be expressed as
$$\hat{\rho}=\int Q_\Gamma(\mathbf{x})\;\hat{D}_\mathbf{x}\hat{\rho}_{-\Gamma}\hat{D}_{\mathbf{x}}^\dag\;d\mathbf{x}\;.$$

Now we are ready to compute the lower bound for the entropy of a state evolved under a dissipative evolution defined as in Eqs.~\eqref{liouville11}, (\ref{liouville}).
First we observe that even though the matrix $\gamma$ entering~(\ref{sigma}) does not necessarily satisfy $\gamma\geq\pm i\Delta$ there exists always a constant $t_1\geq1$ such that $t_1\gamma$  fulfills such inequality, i.e. $t_1 \gamma\geq\pm i\Delta$, the existence of such $t_1$ being ensured by the positivity of $\gamma$. We can hence exploit the generalized Husimi representation~\eqref{rhoQ} associated to the matrix  $\Gamma=t_1\gamma$. For the linearity and the compatibility with translations \eqref{liouvtr} of the evolution \eqref{liouville}, we can take the super-operator $e^{t\mathcal{L}_\gamma}$ that expresses the formal integration of the dissipative process~(\ref{liouville11}) inside the integral:
\begin{equation}
e^{t\mathcal{L}_\gamma}\hat{\rho}=\int Q_{t_1\gamma}(\mathbf{x})\;\hat{D}_\mathbf{x}\left(e^{t\mathcal{L}_\gamma}\hat{\rho}_{-t_1\gamma}\right)\hat{D}_\mathbf{x}^\dag\;d\mathbf{x}\;,
\end{equation}
and since $Q_{t_1\gamma}(\mathbf{x})$ is a probability distribution, the concavity of the von Neumann entropy implies $S\left(e^{t\mathcal{L}_\gamma}\hat{\rho}\right)\geq S\left(e^{t\mathcal{L}_\gamma}\hat{\rho}_{-t_1\gamma}\right)$.
The point now is that for $t>2t_1$, $e^{t\mathcal{L}_\gamma}\hat{\rho}_{-t_1\gamma}$ is a Gaussian state with covariance matrix $(t-t_1)\gamma$, i.e. $e^{t\mathcal{L}_\gamma}\hat{\rho}_{-t_1\gamma}=\hat{\rho}_{(t-t_1)\gamma}$, and for $t\geq2t_1$ it is a proper quantum state. Let $\nu_i,\;i=1,\ldots,n$ be the symplectic eigenvalues of $\gamma$, i.e. the absolute values of the eigenvalues of $\gamma\Delta^{-1}$ \cite{channels}. Remembering that the entropy of the associated Gaussian state is
$$S\left(\hat{\rho}_\gamma\right)=\sum_{i=1}^n h(\nu_i)\;,$$
where
\begin{equation}
h(\nu)=\frac{\nu+1}{2}\ln\frac{\nu+1}{2}-\frac{\nu-1}{2}\ln\frac{\nu-1}{2}\;,
\end{equation}
we have
$$S\left(\hat{\rho}_{(t-t_1)\gamma}\right)=\sum_{i=1}^n h\left((t-t_1)\nu_i\right)\;.$$
Since
$$h(\nu)=\ln\frac{\nu}{2}+1+\mathcal{O}\left(\frac{1}{\nu^2}\right)\qquad\text{for}\;\nu\to\infty\;,$$
we finally get
\begin{eqnarray}
S\left(e^{t\mathcal{L}_\gamma}\hat{\rho}\right)&\geq&\sum_{i=1}^n\ln\frac{e(t-t_1)\nu_i}{2}+\mathcal{O}\left(\frac{1}{t^2}\right)=\nonumber\\
&=&n\ln\frac{et}{2}+\frac{1}{2}\ln\det\gamma+\mathcal{O}\left(\frac{1}{t}\right)\;,\label{lower}
\end{eqnarray}
where in the last step we have used that $\det\gamma=\prod_{i=1}^n\nu_i^2$.

\subsubsection{An upper bound for the entropy}
Given a state $\hat{\rho}$, let $\hat{\rho}_G$ be the Gaussian state with the same first and second moments. It is then possible to prove \cite{gauss} that $S\left(\hat{\rho}_G\right)\geq S\left(\hat{\rho}\right)$.
Since the action of the evolution \eqref{liouville} on first and second moments is completely determined by them (and does not depend on other properties of the state), the Liouvillean $\mathcal{L}_\gamma$ commutes with {\it Gaussianization}, i.e. $\left(e^{t\mathcal{L}_\gamma}\hat{\rho}\right)_G=e^{t\mathcal{L}_\gamma}\left(\hat{\rho}_G\right)$, and we can upper-bound the entropy of the evolved state with the one of the Gaussianized evolved state:
\begin{equation}
S\left(e^{t\mathcal{L}_\gamma}\hat{\rho}\right)\leq S\left(\left(e^{t\mathcal{L}_\gamma}\hat{\rho}\right)_G\right)=S\left(e^{t\mathcal{L}_\gamma}\left(\hat{\rho}_G\right)\right)\;.
\end{equation}
From Eq.~(\ref{sigma}) we know that if $\sigma$ is the covariance matrix of $\hat{\rho}$, the one of $e^{t\mathcal{L}_\gamma}\hat{\rho}$ is given by $\sigma+t\gamma$. Since the entropy does not depend on first moments, we have to compute the asymptotic behaviour of $S(\hat{\rho}_{\sigma+t\gamma})$. Let $t_2>0$ be such that $\sigma\leq t_2\gamma$.
As $\gamma>0$, such $t_2$ always exists: let $\lambda^\downarrow_1$ be the biggest eigenvalue of $\sigma$, and $\mu^\uparrow_1>0$ the smallest one of $\gamma$. Then $\sigma\leq\lambda^\downarrow_1\openone_{2n}\leq\frac{\lambda^\downarrow_1}{\mu^\uparrow_1}\gamma$, so that $t=\frac{\lambda^\downarrow_1}{\mu^\uparrow_1}$ does the job. Now we remind that given two covariance matrices $\sigma'\leq\sigma''$, the Gaussian state $\hat{\rho}_{\sigma''}$ can be obtained applying an additive noise channel to $\hat{\rho}_{\sigma'}$. Since such channel is unital, it always increases the entropy, so  we have $S(\hat{\rho}_{\sigma'})\leq S(\hat{\rho}_{\sigma''})$. Applying this to $\sigma+t\gamma\leq(t_2+t)\gamma$, we get
\begin{eqnarray}
S(\hat{\rho}_{\sigma+t\gamma})&\leq& S\left(\hat{\rho}_{(t_2+t)\gamma}\right)=\sum_{i=1}^n h\left((t_2+t)\nu_i\right)=\nonumber\\
&=&n\ln\frac{et}{2}+\frac{1}{2}\ln\det\gamma+\mathcal{O}\left(\frac{1}{t}\right)\;,\label{upper}
\end{eqnarray}
where in the last step we have used that $\det\gamma=\prod_{i=1}^n\nu_i^2$.

\subsubsection{Scaling of the rate}
Putting together \eqref{lower} and \eqref{upper}, we get
\begin{equation}
e^{\frac{1}{n}S\left(e^{t\mathcal{L}_\gamma}\hat{\rho}\right)}=\left(\det\gamma\right)^\frac{1}{2n}\frac{et}{2}+\mathcal{O}\left(1\right)\;.
\end{equation}
From section \ref{evol} we can see that for our evolutions if $M_\alpha$ is invertible $\det\gamma_\alpha=1$, so $$e^{\frac{1}{n}S_\alpha(\tau)}=\frac{e}{2}t_\alpha(\tau)+\mathcal{O}\left(1\right)$$
and similarly
$$e^{\frac{1}{n}S_Y(\tau)}=\frac{e}{2}t_Y(\tau)+\mathcal{O}\left(1\right)\;.$$
Replacing this into the left-hand-side of Eq.~(\ref{EPIrate}), and remembering that if $M_\alpha$ is not invertible, then $\lambda_\alpha=0$ and the corresponding terms vanish,
from  ~(\ref{div}) and \eqref{tY}  it easily follows that such quantity tends to $1$ in the $\tau\rightarrow \infty$ limit.

\section{Conclusions}
The multi-mode version of the qEPI~\cite{ks,dPMG} has been proposed and proved. This  inequality, while probably not tight, provides a useful bound on the
entropy production at the output of a multi-mode scattering process where independent collections of incoming, multi-mode, incoming inputs collide to produce a given output channel. Explicit examples of such a process are provided by  broadband bosonic channels  where the single signals are described as pulses propagating along optical fibers or in free space communication~\cite{CAVES}.

\section{Acknowledgements}
This work is partially supported by the EU Collaborative Project TherMiQ (grant agreement 618074).

\end{document}